\documentclass[fleqn,multphys,vecphys]{svmult}

\usepackage{makeidx}   
\usepackage{graphicx}  
\makeindex             
\usepackage{hyperref}  

\newcommand{\iDev}[1]{#1}
\newcommand{\iName}[1]{#1}
\newcommand{\iStreet}[1]{#1}
\newcommand{\iPostcode}[1]{#1}
\newcommand{\iCity}[1]{#1}
\newcommand{\iCountry}[1]{#1}
\newcommand{\email}[1]{\texttt{#1}}

\begin{document}

\title*{Correlation effects on electronic transport 
through dots and wires\footnote{In memory of Xavier
  Barnab\'e-Th\'eriault, who passed away in a tragic accident on August
15, 2004.}}          
\toctitle{Correlation effects on electronic transport 
through dots and wires}        
\titlerunning{Correlation effects on electronic transport through dots and wires}                                

\author{V.~Meden}
\authorrunning{V.~Meden} 

\institute{\iDev{Institut f\"ur Theoretische Physik},                  
\iName{Universit\"at G\"ottingen}, \newline
\iStreet{Friedrich-Hund-Platz 1},
\iPostcode{D-37077}
\iCity{G\"ottingen},
\iCountry{Germany}\newline
\email{meden@theorie.physik.uni-goettingen.de}
}

\maketitle

\begin{abstract}
                We investigate how two-particle interactions affect
the electronic transport through meso- and nanoscopic systems
of two different types: quantum dots with local Coulomb
correlations and quasi one-dimensional quantum wires of 
interacting electrons. A recently developed functional renormalization 
group scheme is used that allows to investigate systems of complex
geometry. Considering simple setups we show that the method 
includes the essential aspects of Luttinger liquid physics 
(one-dimensional wires) as well as of the physics of local 
correlations, with the Kondo effect being an important example. 
For more complex systems of coupled dots and Y-junctions of interacting
wires we find surprising new correlation effects.
\end{abstract}

\section{Introduction}
In recent years the effect of electron correlations on the physics of
meso- and nanoscopic systems has attracted growing interest. This led
to an increasing overlap of the two communities working on mesoscopic
physics and strongly correlated electron systems. In this article we 
focus on electronic transport properties, in particular the linear response 
conductance $G$. As discussed below two very fundamental correlation 
effects should be observable in transport through mesoscopic 
systems of relatively simple structure:  (i) the Kondo 
effect\cite{Hewson} in single quantum dots\cite{Glazman,Ng}  and 
(ii) Luttinger liquid (LL) physics in quasi one-dimensional (1d) 
quantum wires with a single impurity.\cite{KS} 

The progress in nanostructuring techniques makes it now possible to 
design more complex geometries such as double- and triple-dot
systems\cite{Holleitner1,Craig} and junctions of several quasi 1d 
wires.\cite{Terrones} In these systems one can expect to find even more
interesting correlation physics.  In the near future 
complex setups might be used in conventional devices as well as 
for quantum information processing which provides a second reason to
investigate the role of correlations. The theoretical
tools commonly used to study many-body physics in dots and wires are rather
specific to simple setups and cannot directly be
applied to more complex geometries. Thus, there is need for novel
techniques which can properly describe correlations, but are flexible
and simple enough such that they can be used to investigate complex
systems. In this article, we show that an approximation scheme which is
based on the functional renormalization group\cite{Manfred} (fRG) 
provides such a method. We will first apply it to the two simple 
setups mentioned above and show that it contains the essential
physics. We then proceed
and study a system of parallel double-dots and a Y-junction of three
1d quantum wires. In both examples the electron correlations lead to 
surprising new effects.  

To exemplify the importance of correlations 
we first study transport through a quantum dot with spin degenerate
levels. For simplicity we only consider a single level (e.g.~described
by the single
impurity Anderson model\cite{Hewson}) and equal 
couplings to the left and right lead.  The level position
can be moved by a gate voltage $V_g$.
At small $T$ and for 
noninteracting dot electrons $G(V_g)$ shows a Lorentzian resonance 
of unitary height $2 e^2/h$. The full width $2 \Gamma$ of the 
resonance sets an energy scale $\Gamma$ which is associated with 
the strength of the tunneling barriers. Including a Coulomb
interaction $U$ between the 
spin up and down dot electrons the line shape is substantially
altered as can be seen from the exact $T=0$ Bethe ansatz
solution.\cite{Tsvelik} 
For increasing $U/\Gamma$ it is gradually transformed into a 
box-shaped resonance of unitary height with a plateau of width $U$ 
and a sharp decrease of $G$ to the left and right of 
it.\cite{Glazman,Ng,Theo1,Gerland} For gate voltages within the
plateau the dot is half-filled implying a local spin-1/2 degree of 
freedom on the dot. Thus, the Kondo effect\cite{Hewson} 
leads to resonant transport throughout this so-called Kondo
regime. The appearance of the plateau can be understood by studying
the
characteristics of the one-particle spectral function of the dot.
For the present setup $G(V_g)$ is proportional to the spectral weight 
at the chemical potential 
$\mu$.\cite{Meir} For sufficiently large
$U/\Gamma$ the spectral function shows a sharp Kondo resonance
(and additional Hubbard bands at higher energies). Its
width sets an energy scale---the Kondo temperature
$T_K$.\cite{Theo1,Hewson} At half-filling ($V_g=0$) the peak is located at the
chemical potential, but even for gate voltages away
from the particle-hole symmetric point it is pinned at $\mu$ 
and its height barely changes. This holds for $-U/2 < V_g
<U/2$, which explains the appearance of the plateau of width $U$ in
the conductance.  
Experimentally the appearance of Kondo physics in transport
through quantum dots was demonstrated clearly.\cite{Goldhaber,Wiel}

Although local correlations in a quantum dot (i.e.~a zero-dimensional
system) already have a strong effect, the system remains a Fermi
liquid. However, 
the low-energy physics of correlated 1d metals (quantum wires) 
is not described by the Fermi liquid theory. Such systems fall into the  
LL universality class\cite{KS} which is
characterized by power-law scaling of a variety of 
correlation functions and a vanishing quasi-particle weight. 
For spin-rotational invariant interactions and spinless models,
on which we focus here, the exponents of the different correlation 
functions can be parametrized by a single number, the interaction dependent 
LL parameter $K < 1$ (for repulsive interactions; $K=1$ in 
the noninteracting case).  
Instead of being quasi-particles the low lying excitations 
of LLs are collective density excitations. This implies that 
impurities, or more generally inhomogeneities, have a dramatic effect 
on the physical properties of 
LLs.\cite{LutherPeschel,Mattis,ApelRice,Giamarchi}
In the presence of only a single impurity on 
asymptotically small energy scales observables behave as if the  1d system 
was cut in two halfs at the position of the impurity, with open boundary 
conditions at the end points 
(open chain fixed point).\cite{KaneFisher,Furusaki0,Fendley} 
Within a renormalization group approach the impurity increases
from weak to strong.
For a weak impurity and decreasing  energy scale $s$---say
the temperature $T$---the  deviation of the linear conductance $G$ 
from the impurity-free  value first scales as $(s/s_0)^{2(K-1)}$, 
with $K$ being the scaling dimension of the perfect chain fixed point 
and $s_0$ a characteristic energy scale (e.g.~the band width). This 
holds as long as $\left|V_{\rm back}/s_0\right|^2 (s/s_0)^{2(K-1)} \ll 1$, with 
$V_{\rm back}$ being a measure for the strength of 
the $2k_F$ backscattering of the impurity and $k_F$ the Fermi momentum.
For smaller energy scales or larger bare impurity backscattering this 
behavior crosses over to another power-law scaling 
$G(s) \sim (s/s_0)^{2(1/K-1)}$, with the scaling dimension of the open chain 
fixed point $1/K$. This scenario was verified within an effective 
field theoretical model for infinite 
LLs\cite{KaneFisher,Furusaki0,Fendley} as well as finite LLs  
connected to Fermi liquid leads.\cite{FurusakiNagaosa} 
In the latter case the scaling holds as 
long as the contacts are modeled to be ``perfect'', that is free
of any bare and effective single-particle backscattering, and
the  impurity is placed in the bulk of the interacting quantum wire. 
Indications of power-law scaling of $G(T)$ were obtained in
experiments on quasi 1d wires, but the results are 
ambiguous.\cite{experiments} One
reason for this is that experimentally the power-law behavior is 
restricted to less than one order of magnitude and often only achieved 
after a somewhat uncontrolled background subtraction on the data.

\section{The RG method and its application to simple systems} 

The fRG was recently introduced as a powerful new  
tool for studying interacting Fermi systems.\cite{Manfred} 
It provides a systematic
way of resumming competing instabilities and goes
beyond simple perturbation theory even in problems which are not 
plagued by infrared divergences.\cite{Ralf} 
In our applications the dot(s) as well as the interacting quantum
wire(s) will be coupled to noninteracting leads as it is the case 
in systems which can be realized in experiments. Before setting up the
fRG scheme we integrate out the leads.\cite{Tilman}  
In the fRG procedure the noninteracting propagator 
$\mathcal{G}_0$ (now including self-energy contributions from the
leads) is replaced by a propagator 
depending on an infrared cutoff $\Lambda$. Specifically,
we use 
\begin{eqnarray}
\label{cutoffproc}
{\mathcal G}_0^{\Lambda}(i \omega) = \Theta(|\omega|-\Lambda) 
{\mathcal G}_0(i \omega)
\end{eqnarray}
with $\Lambda$ running from $\infty$ down to $0$. 
Using 
${\mathcal G}_0^\Lambda$ in the 
generating functional of the irreducible vertex 
functions and taking the derivative with respect to 
$\Lambda$ one can derive an exact, infinite hierarchy of coupled differential
equations for the vertex functions, such as the self-energy and the
irreducible 2-particle interaction. In 
particular, the flow of the self-energy $\Sigma^\Lambda$ (1-particle
vertex) is determined by $\Sigma^\Lambda$ itself and 
the 2-particle vertex $\Gamma^{\Lambda}$, while the flow of 
$\Gamma^{\Lambda}$ is determined by $\Sigma^\Lambda$, $\Gamma^{\Lambda}$, and 
the flowing 3-particle vertex $\Gamma_3^{\Lambda}$.
The latter could be computed from a flow equation involving
the 4-particle vertex, and so on.
At the end of the fRG flow $\Sigma^{\Lambda=0}$ is the self-energy $\Sigma$ 
of the original, cutoff-free problem we are interested in.\cite{Ralf}
A detailed derivation of the fRG flow equations for a general quantum
many-body problem which only requires a basic knowledge of the
functional integral approach to many-particle
physics and the application of the method for a
simple toy problem are presented in Ref.~\cite{lecturenotes}.   

In practical applications the hierarchy of flow 
equations has to be truncated and $\Sigma^{\Lambda=0}$ only provides 
an approximation for the exact $\Sigma$. As a first approximation 
we here neglect the 3-particle vertex.
The contribution of $\Gamma_3^{\Lambda}$ to $\Gamma^{\Lambda}$ is small
as long as $\Gamma^{\Lambda}$ is small, because 
$\Gamma_3^{\Lambda}$ is initially (at $\Lambda=\infty$) zero 
and is generated only from terms of third order in 
$\Gamma^{\Lambda}$. Furthermore,  $\Gamma^{\Lambda}$ stays small 
for all $\Lambda$ if the bare interaction is not too large. 
Below we will clarify the meaning of ``not-too-large'' 
in the cases of interest. This approximation leads to a closed set of 
equations for $\Gamma^{\Lambda}$  and
$\Sigma^{\Lambda}$.\cite{Sabine}  We here do not give these equations
but instead implement a second approximation: the
frequency-dependent flow of the renormalized 2-particle vertex 
$\Gamma^{\Lambda}$ is replaced by its value at vanishing (external) 
frequencies, such that $\Gamma^{\Lambda}$ remains frequency
independent.
Since the bare interaction is frequency independent, neglecting the
frequency dependence leads to errors only at second order for 
the self-energy, and at third order for the 2-particle vertex at 
zero frequency.
For the approximate flow equations we then obtain
\begin{eqnarray}
 \frac{\partial}{\partial\Lambda} \Sigma^{\Lambda}_{1',1} =
 - \frac{1}{2\pi} \sum_{\omega = \pm \Lambda} \sum_{2,2'} \,
 e^{i\omega 0^+} \, {\mathcal G}^{\Lambda}_{2,2'}(i\omega) \,
 \Gamma^{\Lambda}_{1',2';1,2} 
\label{finalflowsigma}
\end{eqnarray}
and 
\begin{eqnarray}
&& \frac{\partial}{\partial\Lambda} \Gamma^{\Lambda}_{1',2';1,2}   = 
\frac{1}{2\pi} \, 
 \sum_{\omega = \pm\Lambda} \, \sum_{3,3',4,4'} 
 \Big\{ \frac{1}{2} \,  {\mathcal G}^{\Lambda}_{3,3'}(i\omega) \, 
  {\mathcal G}^{\Lambda}_{4,4'}(-i\omega) 
 \Gamma^{\Lambda}_{1',2';3,4} \, \Gamma^{\Lambda}_{3',4';1,2} 
\nonumber \\ && +  {\mathcal G}^{\Lambda}_{3,3'}(i\omega) \,
 {\mathcal G}^{\Lambda}_{4,4'}(i\omega)  
 \left[ - \Gamma^{\Lambda}_{1',4';1,3} \, \Gamma^{\Lambda}_{3',2';4,2} 
        + \Gamma^{\Lambda}_{2',4';1,3} \, \Gamma^{\Lambda}_{3',1';4,2}
 \right] \Big\} \; ,
\label{finalflowgamma}
\end{eqnarray}
where the lower indexes $1$, $2$, etc.\  stand for the 
single-particle quantum numbers and
\begin{eqnarray}
\label{Glambdadef}
   {\mathcal G}^{\Lambda}(i\omega) = 
 \left[  {\mathcal G}_0^{-1}(i\omega) - \Sigma^{\Lambda} \right]^{-1} \; .
\end{eqnarray}
At the initial cutoff $\Lambda=\infty$ the flowing 2-particle vertex
$\Gamma^{\Lambda}_{1',2';1,2}$  is given by the antisymmetrized
interaction and the self-energy by the single-particle terms of the
Hamiltonian not included in ${\mathcal G}_0$ (e.g.~impurities). 

\subsection{The single-level quantum dot}
Now the set of flow equations can be used to study the
two-lead single impurity Anderson model.\cite{Hewson,dotsystempaper} 
After integrating out the leads the only relevant single-particle
quantum number is the spin $\sigma$ of the dot electrons. 
If we take the level energies to
be $\varepsilon_{\uparrow}=V_g-{\mathcal H}/2$ and
$\varepsilon_{\downarrow}=V_g+{\mathcal H}/2$,  where ${\mathcal H}$ 
denotes a magnetic field which lifts the spin-degeneracy, the 
projected noninteracting propagator in the so-called wide band limit 
is given by\cite{Hewson}
\begin{eqnarray}
\label{singledotprop0}
  {\mathcal G}_{0,\sigma} (i
  \omega) = \frac{1}{i \omega -(V_g + \sigma {\mathcal H}/2) + 
i \Gamma \,\mbox{sgn}(\omega)}  \; ,
\end{eqnarray}
where on the right-hand side 
$\sigma = \,\uparrow\, = +1$ and  $\sigma = \,\downarrow\, = -1$. 

During the fRG flow $\Sigma^\Lambda$ remains frequency independent and 
$V_\sigma^\Lambda=V_g+ \sigma {\mathcal H}/2+\Sigma_{\sigma}^\Lambda$ 
can be interpreted as an effective, flowing level position, whose 
flow equation reads
\begin{eqnarray}
\label{sigmasingledot}
\frac{\partial }{\partial \Lambda}V_\sigma^\Lambda & = & - 
\frac{U^\Lambda}{2 \pi} \, \sum_{\omega=\pm \Lambda} 
 {\mathcal G}^{\Lambda}_{\bar \sigma}(i
\omega) \nonumber \\
& = & 
\frac{U^\Lambda \, V_{\bar\sigma}^\Lambda/\pi}{(\Lambda+ \Gamma)^2
+(V_{\bar \sigma}^\Lambda)^2} \, ,
\end{eqnarray}   
with the initial condition $V_\sigma^{\Lambda=\infty}=V_g+\sigma {\mathcal H}/2$ 
and $\bar \sigma$ denoting the complement of $\sigma$. 
The cutoff-dependent propagator $ {\mathcal G}^{\Lambda}_{\sigma}(i
\omega) $ follows from $  {\mathcal G}_{0,\sigma} (i \omega)$ by replacing 
$V_g + \sigma {\mathcal H} /2 \to V_\sigma^\Lambda$. Using symmetries, 
the flow of the 2-particle vertex can be reduced to a single 
equation for the effective interaction between spin up and spin down 
electrons 
\begin{eqnarray}
\label{Usingledot}
\frac{\partial }{\partial \Lambda}U^\Lambda  
 & \! \! \! \! =  \!\! \!\!  &  \frac{(U^\Lambda)^2}{2 \pi} \!\!  
\sum_{\omega=\pm \Lambda} \! \! \left[ 
\tilde  {\mathcal G}^{\Lambda}_{\uparrow}(i \omega) \,
\tilde   {\mathcal G}^{\Lambda}_{\downarrow}(-i \omega)
+\tilde  {\mathcal G}^{\Lambda}_{\uparrow}(i \omega) \,
\tilde   {\mathcal G}^{\Lambda}_{\downarrow}(i \omega)
  \right]
\nonumber \\ 
& \!=  \! &  \frac{2\, \left(U^\Lambda\right)^2\, 
V_\uparrow^\Lambda \, V_\downarrow^\Lambda/\pi}{\left[ 
(\Lambda+ \Gamma)^2
+(V_{\uparrow}^\Lambda)^2 \right]\left[ 
(\Lambda+ \Gamma)^2
+(V_{\downarrow}^\Lambda)^2 \right]} \; ,
\end{eqnarray}   
with $U^\Lambda= \Gamma^\Lambda_{\sigma,\bar \sigma;\sigma, \bar
  \sigma}$ and the initial condition $U^{\Lambda=\infty} = U$. 

\begin{figure}
\begin{center}
\includegraphics[width=0.6\textwidth,clip]{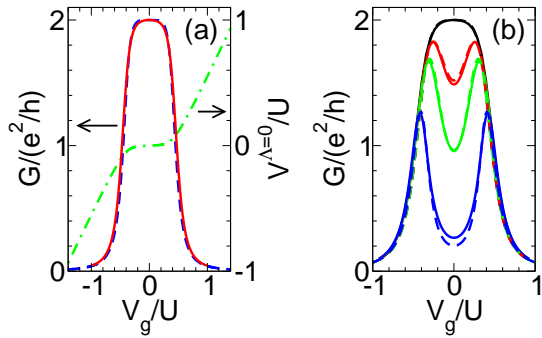}
    \caption{(a) Gate Voltage $V_g$ dependence of the effective 
 level position $V^{\Lambda=0}$ (dashed-dotted
 line) and the linear conductance $G$ (solid line: Bethe ansatz of
 Ref.~\cite{Gerland}; dashed line: fRG) of a single level quantum dot
  for $U/\Gamma=4 \pi$, ${\mathcal H}=0$. (b) $G(V_g)$ 
  for $U/\Gamma=3\pi$ and  ${\mathcal H}=0$, $0.5 T_K$, $T_K$, 
  and $5T_K$, from top to bottom. Here 
  $T_K=0.116 \Gamma$. Solid line: NRG data 
  from Ref.~\cite{Theo2}. Dashed line: fRG
  approximation.}
      \label{medefig1}
\end{center}
\end{figure}

Within our approximation the dot spectral function at the end of the fRG 
flow is given by  
\begin{eqnarray}
\label{dotspecfu}
\rho_\sigma( \omega) = \frac{1}{\pi} \; 
\frac{\Gamma}{\left( \omega - V_\sigma \right)^2 + \Gamma^2} \; ,
\end{eqnarray}
with $V_\sigma = V^{\Lambda=0}_\sigma$, that is a Lorentzian of full 
width $2\Gamma$ and height $1/(\pi \Gamma)$ centered around
$V_\sigma$.  We first consider ${\mathcal H}=0$. Although this 
spectral function neither shows the narrow Kondo resonance nor 
the Hubbard bands, for $U/\Gamma \gg 1$ the pinning  of the 
spectral weight at the
chemical potential is included in our approximation. We thus 
reproduce the  line shape of the conductance quantitatively up 
to very large $U/\Gamma$. This is shown in Fig.~\ref{medefig1}(a) where
we plot $V^{\Lambda=0}$ as a function of $V_g$ (dashed-dotted line) as 
well as $G(V_g)$ (dashed line) for $U/\Gamma=4\pi$. For comparison 
the exact result  for $G(V_g)$ obtained by Bethe 
ansatz\cite{Tsvelik,Gerland}  is shown as the solid line. 
Neglecting the flow of the two-particle
vertex the exponential pinning of spectral weight can even be shown
analytically.\cite{dotsystempaper} 

Even without the sharp Kondo resonance in the spectral function, which 
is usually used to define $T_K$, our approximation contains the 
Kondo temperature. In Fig.~\ref{medefig1}(b) we show a 
comparison of fRG data (dashed lines) and high precision numerical 
renormalization group (NRG) data\cite{Theo2} for $G(V_g)$ and different 
${\mathcal H}$. The $T_K$ given in the caption was obtained from the width 
of the Kondo resonance at $V_g=0$ using NRG. Within NRG at ${\mathcal H}=T_K$ 
(third solid curve from top) $G(V_g=0)=e^2/h$. This exemplifies
that $T_K$ can equally be defined as the magnetic field which is necessary to 
suppress $G$ down to $e^2/h$. This criterion can
be used for the fRG data and the excellent agreement of the fRG curves and the NRG 
results in   Fig.~\ref{medefig1}(b) shows that our approximation (in contrast 
to other simple approximation schemes) indeed contains $T_K$. 
 
We note that using a fRG based truncation 
scheme in which the full frequency dependence of the 2-particle 
vertex is kept (leading to a frequency dependent self-energy) it 
was shown that one can also reproduce the Kondo resonance and 
Hubbard bands of the spectral function, however with 
a much higher numerical effort.\cite{Ralf}    

\subsection{A quantum wire with a single impurity}

\begin{figure}[t]
\begin{center}
\includegraphics[width=0.5\textwidth,clip]{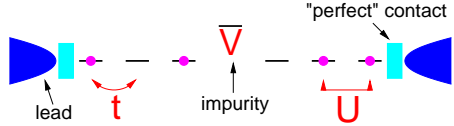}
    \caption{Sketch of the model for an interacting quantum wire with
      nearest-neighbor hopping $t$, nearest-neighbor interaction $U$,
      a local impurity of strength $V$, and $N$ lattice sites. The
      wire is connected to noninteracting leads by  ``perfect'' contacts.}
      \label{medefig2}
\end{center}
\end{figure}

Next we show that the fRG based approximation scheme is also able to produce 
power-law scaling of the conductance, which is 
characteristic for inhomogeneous LLs. 
This requires fairly long chains of interacting electrons---say 
lattice systems 
with $N=10^4$ sites corresponding to a length in the $\mu$m range. Therefore 
the fRG flow equations have to be simplified further. For a spinless 
lattice model of the interacting chain with nearest-neighbor hopping $t$ and 
a nearest-neighbor interaction $U$ (see Fig.~\ref{medefig2})
we achieve this by parameterizing the $\Lambda$ dependent 2-particle 
vertex by a flowing nearest-neighbor interaction $U^\Lambda$.
Then $\Sigma^\Lambda$ is a tridiagonal matrix in the real
space basis of Wannier states
and large systems can be treated.\cite{Sabine}  A similar 
approximation has also been implemented for the (spinful) extended Hubbard 
model.\cite{Sabine2} 
For the spinless model the LL parameter $K$  is known exactly from
Bethe ansatz.\cite{Haldane} Besides $U$ it also depends on the filling
$n$. To compute $G(T)$ one has to generalize 
the fRG to $T>0$. This is described in Refs.~\cite{Tilman} and \cite{Sabine2}. 
We obtain the conductance from the one-particle Green function 
(and thus the self-energy) using a generalized
Landauer-B\"uttiker  
relation.\cite{Tilman} Within our approximation current vertex corrections vanish 
(because $\Sigma^{\Lambda=0}$ is real) and using this relation does
not require any additional approximations.\cite{Tilman} 
As we are only interested in the effect of a single impurity 
we model the contacts between the leads and the interacting wire 
to be ``perfect'', that means for $V=0$ (see Fig.~\ref{medefig2}) the $T=0$ 
conductance through the wire is $e^2/h$. This requires that the
interaction is turned on and off smoothly close to the contacts.\cite{Tilman}

\begin{figure}[t]
\begin{center}
\includegraphics[width=0.63\textwidth,clip]{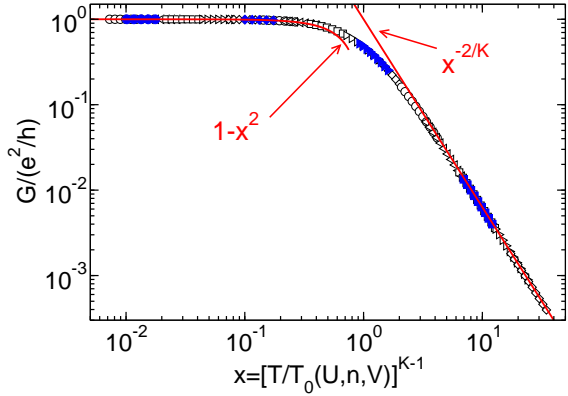}
    \caption{One-parameter scaling of $G(T)$ for a quantum wire with
      $N=10^4$ lattice sites and a single impurity. For details see the text.}
      \label{medefig3}
\end{center}
\end{figure}

The power-law scaling of $G(T)$ close to the perfect and open chain
fixed points can very elegantly be shown in a single plot using a 
one-parameter scaling ansatz. Plotting $G(T)$ as a function 
of $T/T_0$, with an 
appropriately chosen nonuniversal scale $T_0(U,n,V)$, all data obtained 
for different $V$, $U$, and $n$---with the restriction that $K(U,n)$
is fixed---collapse on a single $K$ dependent 
curve.\cite{KaneFisher,Fendley} This is shown in Fig.~\ref{medefig3}. 
The open symbols were obtained for $n=1/2$, $U=0.5$
while the filled ones were computed for $n=1/4$, $U=0.851$, both
parameter sets leading to $K=0.85$. Different symbols stand for
different $V$. For convenience $(T/T_0)^{K-1}$ is
chosen as the $x$-axis such that $G/(e^2/h)$ for $x \ll 1$ scales 
as $1-x^2$ and for $x\gg 1$ as $x^{-2/K}$. 
Within our approximation the scaling exponents are
correct to leading order in $U$.\cite{Sabine,Tilman,Sabine2}  
During the flow the self-energy remains frequency 
independent and $\Sigma^{\Lambda=0}$  can be interpreted as an 
effective impurity potential. Scattering off  
the spatially long-ranged 
oscillations of the self-energy generated during the fRG
flow leads to the observed power-law behavior.\cite{Sabine,Tilman}

We note that in the limits of weak and strong impurities the scaling
can be shown analytically with our fRG scheme.\cite{VM}
Up to now the fRG is the only method which allows to study the entire
crossover from weak to strong impurities for microscopic lattice
models. Studying such models is important as similar to experimental
systems, they contain energy scales that set upper and lower bounds
for scaling. They are absent if field theoretical models are used.

\section{Correlation effects in more complex geometries}
After showing that our method contains the essential physics in the
case of locally correlated systems as well as for inhomogeneous LLs we
proceed and study more complex setups.

\subsection{Novel resonances in a parallel double-dot system} 
\begin{figure}[t]
\parbox{3.9cm}{\includegraphics[width=0.3\textwidth,clip]{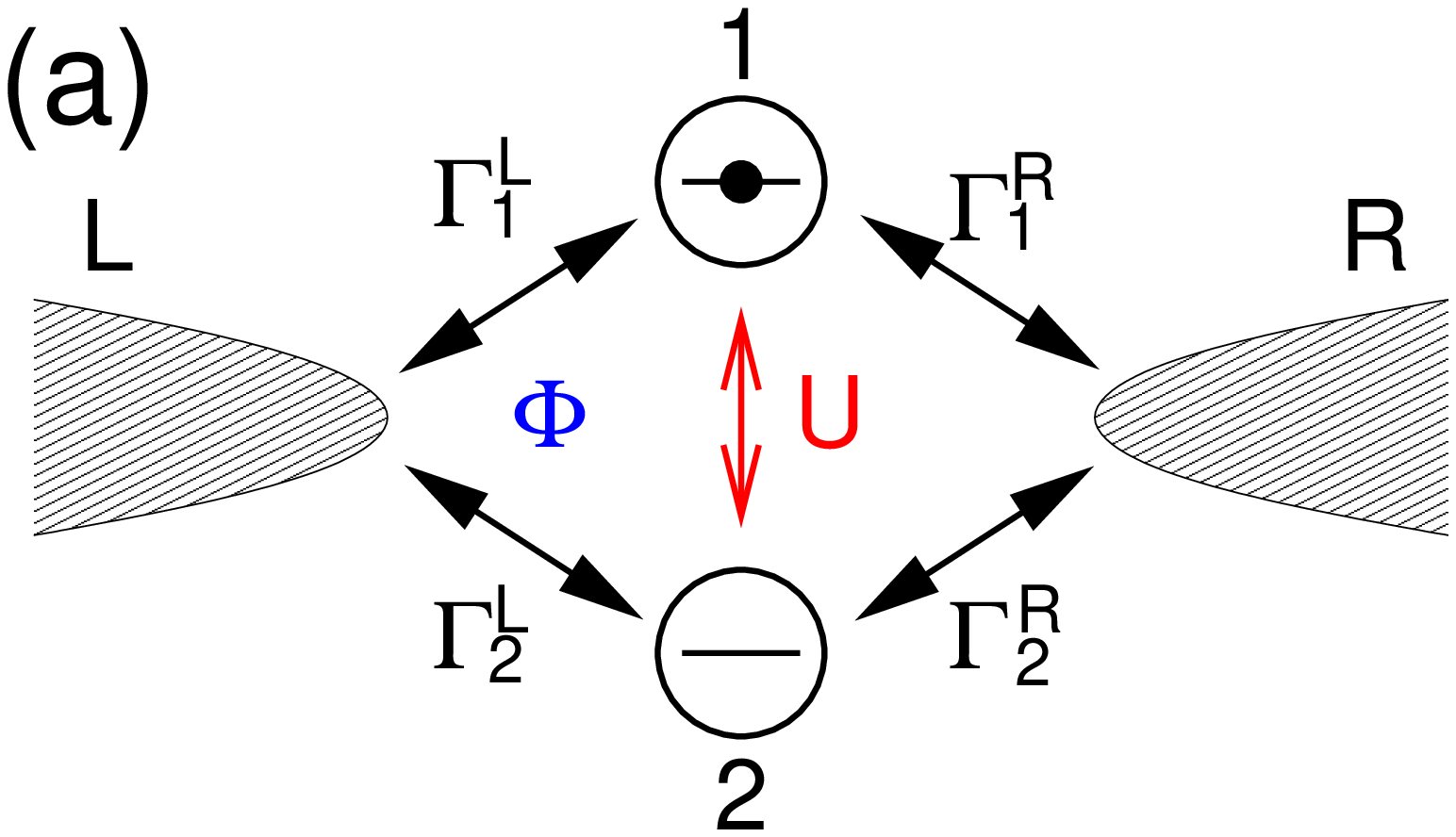}}
\parbox{5cm}{\includegraphics[width=0.4\textwidth,clip]{medefig4b.eps}}
\parbox{4cm}{\includegraphics[width=0.22\textwidth,clip]{medefig4c.eps}}
    \caption{(a) Sketch of the double-dot setup.  (b) Generic results 
for $G(V_g)/(e^2/h)$ at different $U$ obtained 
from the fRG with $\Gamma_1^L=0.27 \Gamma$, $\Gamma_1^R=0.33 \Gamma$, 
$\Gamma_2^L= 0.16 \Gamma$, $\Gamma_2^R=0.24 \Gamma$, and $\phi=\pi$. 
The two novel 
correlation induced resonances are visible in the lower 
panels (large $U$), near $V_g=0$. (c) $U$ dependence of the position
$V_{\rm CIR}$ of the novel resonances (line: fRG; circles: NRG).}
      \label{medefig4}
\end{figure}

As a first example we consider a parallel single-level double-dot
system as sketched in Fig.~\ref{medefig4}(a). The dots are coupled to
common leads by tunneling barriers and the electrons interact by an
inter-dot interaction $U$. The ring geometry is pierced by a magnetic
flux $\phi$. We here neglect the spin 
and thus suppress the spin Kondo effect. Experimentally, the 
contribution of spin physics may be excluded by applying a
strong magnetic field.
We focus on temperature $T=0$. For this setup the flow equations
(\ref{finalflowsigma}) and (\ref{finalflowgamma}) can directly be
applied. $G$ is computed from a generalized Landauer-B\"uttiker 
relation.\cite{dotsystempaper}

Fig.~\ref{medefig4}(b) shows the evolution of $G(V_g)$ 
for degenerate levels $\varepsilon_j=V_g$, generic parameters\cite{doubledotpaper}
$\Gamma_j^l$, $\phi$, and increasing $U$. 
At $U=0$, $G(V_g)$ is a Lorentzian at large $|V_g|$, while it shows a
dip at $V_g=0$. 
With increasing $U$ the height of the two peaks resulting from the dip 
at $V_g=0$ increases and the maximum flattens. At a critical 
$U=U_c(\{\Gamma_j^l\},\phi)$ each peak splits into two. 
For the present example the fRG approximation is 
$U_c/\Gamma \approx 4.69$, with $\Gamma$ being 
the sum over all $\Gamma_j^l$. Further increasing $U$ the two outer most 
peaks move towards larger $|V_g|$ and become the Coulomb blockade 
peaks located at $V_g \approx \pm U/2$. 
The other two peaks at $\pm V_{\rm CIR}$, where $V_{\rm CIR} >0$ 
decreases with increasing $U$, are novel
resonances following from the interplay of quantum interference and
correlations.  For $U> U_c$ the height of all four peaks is equal 
and does not change with $U$. 
Note that in contrast to Ref.~\cite{doubledotpaper} here the flow
of the 2-particle vertex is included which leads to improved 
results. The above scenario was confirmed using
NRG.\cite{doubledotpaper} Furthermore, the appearance of the novel
resonances is robust: It appears for almost arbitrary combinations 
of the four tunnel couplings and the magnetic flux, also remains 
visible for a small detuning of the dot level
energies,\cite{doubledotpaper} and small temperatures.  In
Fig.~\ref{medefig4}(c) we show the dependence of $V_{\rm CIR}$ on
$U$ obtained from fRG (line) and NRG (circles) on a linear-log scale.  
Both curves follow a straight line which shows that 
\begin{eqnarray}
V_{\rm CIR}/\Gamma \propto \exp{\left[-
    C\left(\left\{\Gamma_j^l\right\},\phi\right) U/\Gamma\right]} \, ,   
\label{centralresult}
\end{eqnarray}  
with $C>0$. The fRG apparently underestimates $C$. Within the fRG for 
a certain class of $\{\Gamma_j^l\},\phi$ the appearance of an energy 
scale depending exponentially on $U/\Gamma$ can be shown
analytically.\cite{doubledotpaper} We note that the above correlation
effect is unrelated to spin and orbital Kondo physics.

Double-dot geometries that could form the basis to verify the above  
predictions have been experimentally realized in 
Ref.~\cite{Holleitner1}.

\subsection{Novel fixed points in a  Y-junction of three quantum wires} 

\begin{figure}[t]
\parbox{3.9cm}{\includegraphics[width=0.25\textwidth,clip]{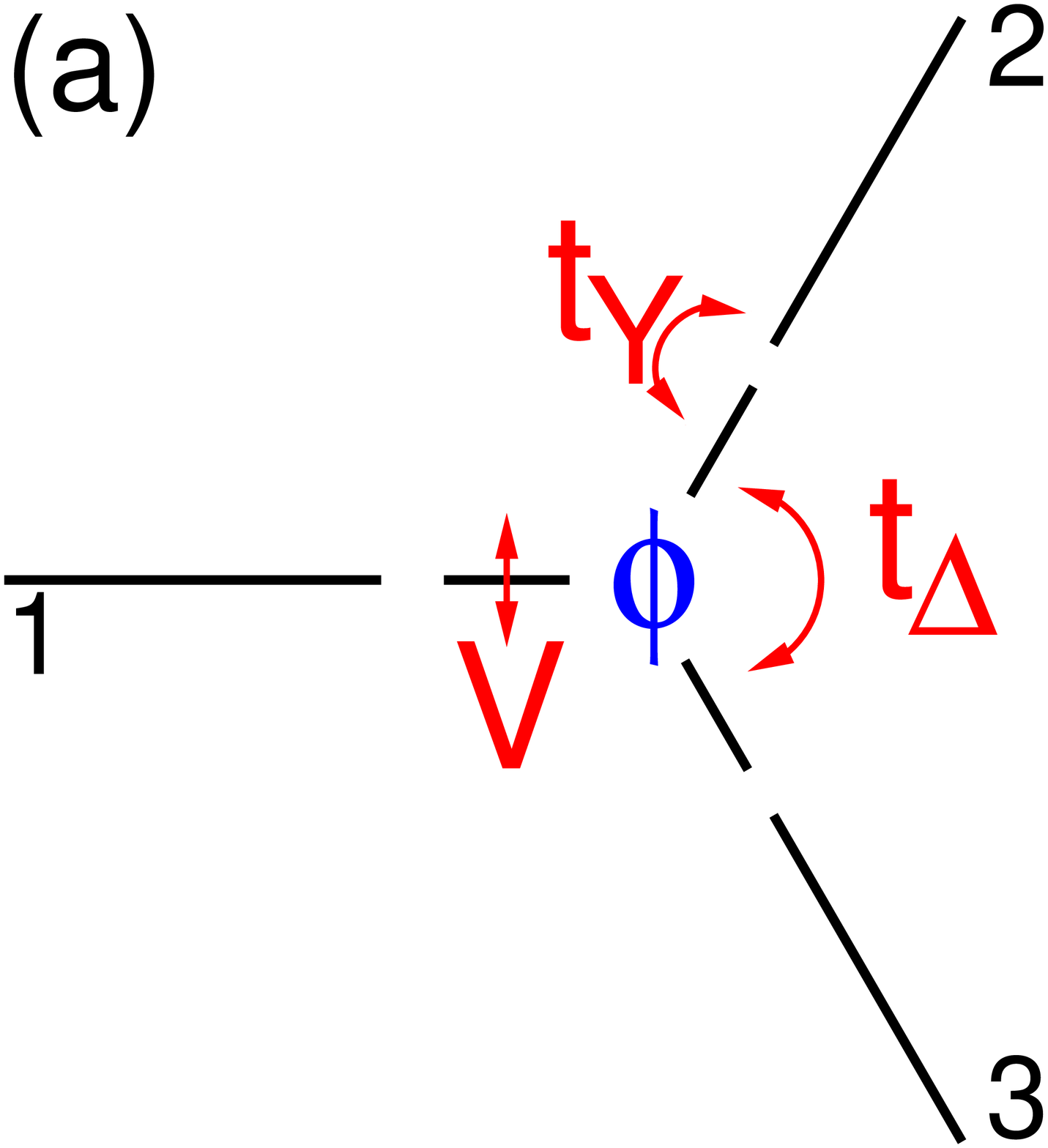}}\hspace{1.2cm}
\parbox{10cm}{\includegraphics[width=0.57\textwidth,clip]{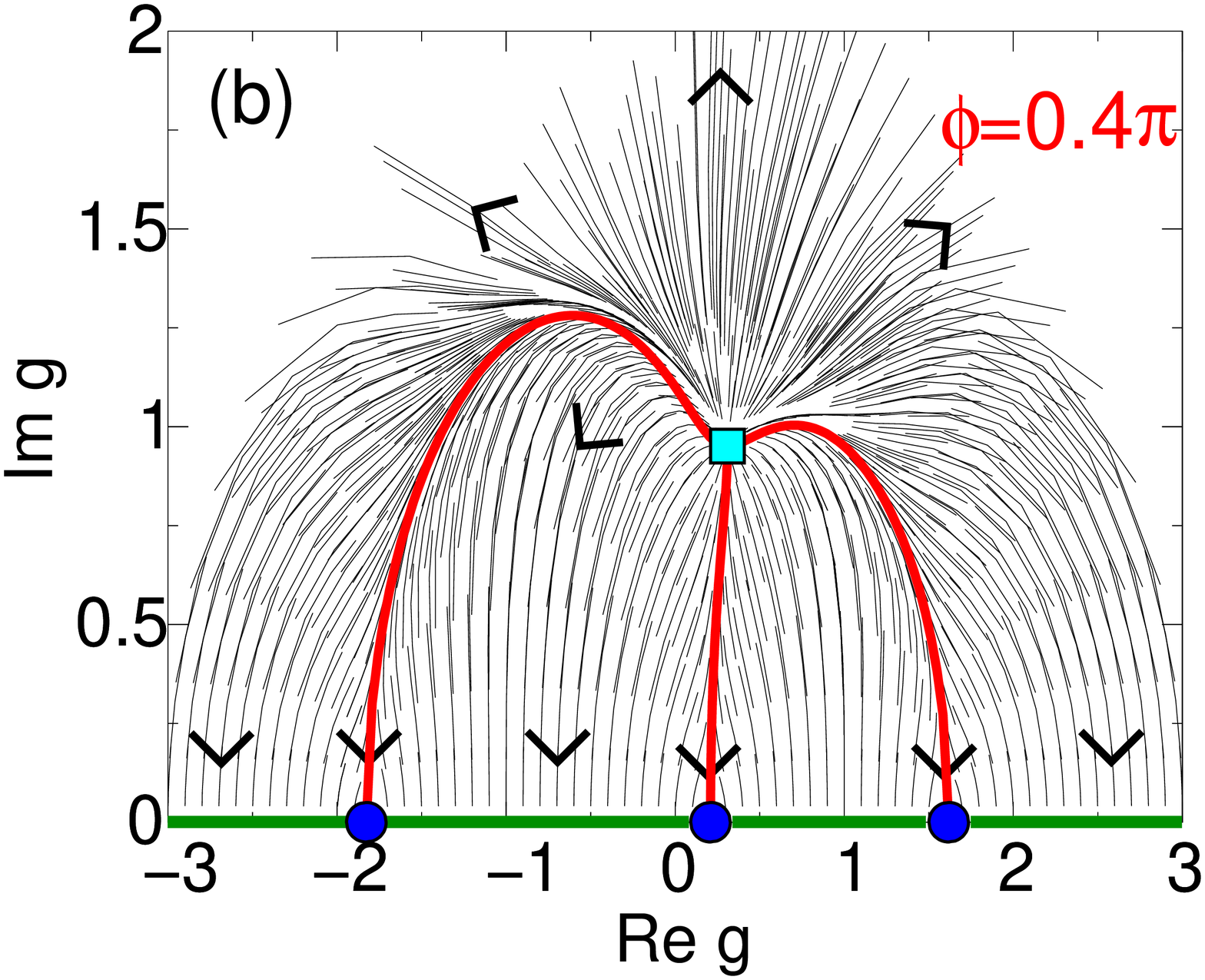}}
    \caption{(a) Setup of the Y-junction of three interacting quantum
      wires.  (b) $U>0$ fRG flow within the complex 
        $g$ plane. For each
      point in the $g$ plane the conductance follows using
      Eq.~(\ref{trans}). The symbols indicate fixed points. The
      condition $\mbox{Im}\, g=0$ defines a line of fixed points. 
     For further details see the text.}
      \label{medefig5}
\end{figure}

As a second example for the application of the fRG to more complex
geometries we study the conductance through a Y-junction of 
interacting quantum wires. The three wires are assumed to be equal 
and described  by the model of
spinless fermions with nearest-neighbor hopping $t$ and  nearest-neighbor 
interaction $U$ as already used above. Each  wire is connected 
to noninteracting leads by ``perfect'' contacts. The junction is
assumed to be symmetric and characterized by three parameters $t_Y$,
$t_{\Delta}$, and $V$ as sketched in Fig.~\ref{medefig5}(a). 
The ring structure is furthermore pierced by a magnetic flux $\phi$
and for $\phi \neq n \pi$, with an integer $n$, time-reversal symmetry 
is broken. For generic junction parameters in the non-interacting case 
this leads to an asymmetry of the conductance from wire $\nu$ to wire 
$\nu'$  (with $\nu,\nu'=1,2,3$) and vice versa $G_{\nu,\nu'} 
\neq G_{\nu',\nu}$ and the breaking of time-reversal symmetry is 
indicated by the conductance. This can be seen using single-particle 
scattering theory. $G_{\nu',\nu}$ can be expressed in terms of the 
single complex and junction parameter dependent number $g=(-V-t_Y^2 \,
\tilde \mathcal G_{1,1})/|t_{\triangle}|$ as
\begin{eqnarray} 
\label{trans}
G_{\nu,\nu'} = \frac{4 \left( \mbox{Im} \, g \right)^2 
\left| e^{-i \phi} - g \right|^2}{\left|
  g^3-3 g+ 2 \cos{\phi} \right|^2} \; ,
\end{eqnarray}
with $\nu,\nu'$ in cyclic order.\cite{yjunctionfluxpaper}
$G_{\nu',\nu}$ follows by replacing 
$\phi \to - \phi$. Here $\tilde \mathcal G$ 
denotes the (wire index independent) single 
particle Green function of one of the semi-infinite wires obtained
after setting $t_Y=0$ and taking the energy 
$\varepsilon +i0$ with $\varepsilon \to 0$. The index $1,1$ stands 
for its diagonal matrix element taken at the first site. At $U=0$
it is given by $\tilde \mathcal G_{1,1}=i/t$.\cite{Tilman} 

At $T=0$ Eq.~(\ref{trans}) also holds for $U>0$. In this case 
$\tilde \mathcal G_{1,1}(\Sigma)$ becomes self-energy and thus $U$ 
dependent. As for the single interacting wire with one 
impurity $\Sigma$ can approximately be computed using the 
fRG. As scaling 
variable we this time use an energy scale $\delta_N=\pi v_F/N$, 
with $v_F$ being the $U=0$ Fermi velocity, set by the length of 
the three interacting wires. As $\Sigma^{\Lambda=0}$ becomes
$\delta_N$ dependent, also $g$ and thus $G_{\nu,\nu'}$ will depend on
the length of the interacting wires. Increasing $N$ 
(that is lowering the infrared cutoff $\delta_N$) for fixed $U,\phi
\neq n\pi, t_Y, t_{\Delta},V$ then leads to a flow in the complex $g$
plane and $G_{\nu,\nu'}$ changes according to Eq.~(\ref{trans}).
The flow for $U=1$ and
$\phi=0.4 \pi$ is shown in Fig.~\ref{medefig5}(b) for a variety 
of different $t_Y, t_{\Delta},V$.  Each line stands for a fixed set of
the junction parameters and varying $\delta_N$. The direction of the
flow is indicated by the arrows. 

The square indicates a fixed point at which the asymmetry is maximal,
that is $G_{1,2}=e^2/h$ and $G_{2,1}=0$. For $U>0$ it is 
unstable.\cite{Claudio,yjunctionfluxpaper} 
For generic junction parameters the flow is directed towards the line 
of fixed points defined by the $\mbox{Im}\, g=0$ line. As is apparent from 
Eq.~(\ref{trans}) on this line and for generic $\mbox{Re}\, g$, both 
$G_{1,2}$ and $G_{2,1}$ vanish. This line of fixed points is the
analogue of the open chain fixed point obtained for one impurity
in a single wire. Analyzing the scaling close to the line of fixed
points in more detail we find that for $\delta_N\to 0$,
$|G_{1,2}-G_{2,1}|/(G_{1,2}+G_{2,1}) \to 0$, i.e.~$G_{1,2}$ and
$G_{2,1}$  become equal faster than they go to zero. Even more
surprising behavior is found if the bare parameters are tuned such that
the flow starts on the thick line. In that case it leads to one
of the three fixed points indicated by the circles. On these fixed
points we find $G_{1,2} = G_{2,1}=(4/9) (e^2/h)$, which is the
conductance maximally allowed by the unitarity of the scattering
matrix for a symmetric Y-junction of noninteracting wires. Combined,
these two observations show that due to the interaction on very low
energy scales (i.e.~for long interacting wires and at low
temperatures) the conductance no longer indicates  a broken
time-reversal symmetry. In that sense the electron
correlations restore the time-reversal 
symmetry.\cite{yjunctionfluxpaper}   

Associated to the novel $(4/9) (e^2/h)$ fixed points is a novel
scaling dimension that should show up in e.g.~the finite temperature
corrections of $G$ with respect to the fixed point 
conductance.\cite{yjunctionfluxpaper}

\section{Summary}
In the present article we have shown that electron 
correlations in nano- and mesoscopic systems can lead to a variety 
of surprising effects. While for simple geometries established methods
are available to investigate important aspects of the many-body problem,
they cannot directly be applied to more complex setups, such as
systems of locally correlated quantum dots and junctions of
interacting quantum wires. To investigate such systems, which in the
near future will shift towards the focus of experimentally activities, 
we introduced a reliable, simple to implement, and numerically very
fast approximation scheme which is based on the functional
renormalization group. In certain limiting cases it can also be 
used  to obtain analytical results.

The author would like to thank T.~Costi and J.~von Delft for providing
their NRG and Bethe ansatz data, S.~Andergassen, T.~Enss, 
C.~Karrasch, F.~Marquardt, W.~Metzner, U.~Schollw\"ock, 
K.~Sch\"onhammer, and A.~Sedeki for collaboration on the issues
presented, and P.~W\"achter for useful comments on the manuscript. 
This work was supported by the Deutsche
Forschungsgemeinschaft (SFB 602).

\end{document}